\documentclass[12pt,a4paper]{article}
\usepackage{amsmath, amssymb}
\newcommand{\be}{\begin{equation}}
\newcommand{\ee}{\end{equation}}
\newcommand{\bear}{\begin{eqnarray}}
\newcommand{\eear}{\end{eqnarray}}

\title{\bf Electrostatic time dilation and redshift}
\author{{Murat \"Ozer\footnote{\texttt{mhozer@yildiz.edu.tr}}  }
\bigskip\\
Y{\i}ld{\i}z Teknik \"Universitesi, Fizik B\"ol\"um\"u, 34220 Esenler, {\.I}stanbul, Turkey\\
$\&$\\
TUSA\c{S}, YT\"U Y{\i}ld{\i}z Teknopark, 34220 Esenler, {\.I}stanbul, Turkey\\}
\date{}
\begin{document}
\maketitle
\begin{abstract}

\noindent We first present the salient features of the gravitational time dilation and redshift effects in two ways; by considering the oscillation frequencies/rates of clocks at different heights/potentials and by considering the photons emitted by these clocks such as atoms/nuclei. We then
point out to the extension of these gravitational effects to static electricity along with two experiments performed in the '30s with null results of the electrostatic redshift. We show that the absence of this redshift is a consequence of the conservation of electric charge. We discuss the electrical time dilation and redshift effects in detail and argue that the electrostatic time dilation in an electric field must be a fact of Nature. We then present a general relativistic scheme that explains this effect.  We also introduce an electrical equivalence principle analogous to the gravitational one and demonstrate how to obtain the electrostatic time dilation by this principle. We emphasize the importance of ionic atomic clocks to measure this effect whose confirmation would support the general relativistic scheme presented. We finally go over an attempt in the literature to explain the impossibility of the experimental observation of the electrostatic redshift due to its smallness by employing the Reissner - Nordstr\"om metric in general relativity. We argue that the $Q^{\,2}$ - term in this metric is due to the minuscule contribution of the energy of the electric field of the central body to its gravitational field. Thus being gravitational, this metric cannot be used to calculate the amount of the alleged electrostatic redshift.\\
\end{abstract}


%\symbolfootnote[0]{E-mail: mhozer@yildiz.edu.tr; murat.h.ozer@gmail.com}

\section{Introduction}

That time passes differently at different heights or potentials in a gravitational field is called  the gravitational time dilation. The time intervals between two events measured by observers located at different altitudes from a gravitational source, a large mass, happen to be different. Time passes faster, in other words, the rate of a clock, namely its oscillation frequency, increases as it gets located farther from a gravitating source. The effect was first predicted by Einstein \cite{eins,schw} and was experimentally verified indirectly by means of the M\"osbauer effect in \cite{p-r,p-s} and directly  in \cite{hafele,alley} by readings of the airborne and earthbound atomic clocks, and in \cite{vessot1,vessot2} by comparing the frequencies of microwave signals from hydrogen maser clocks in a rocket at a high altitude and at an Earth station. The precision of these experiments were improved recently by measuring the frequencies of the onboard hydrogen maser clocks in Galileo Satellites of the European Space Agency \cite{galileo-1, galileo-2}. It was reported in \cite{chou} that time dilation due to a change in height less than a meter could be detected by comparing two optical clocks based on $^{27}Al^+$ ions.

A directly related concept is the redshift of light in a gravitational field \cite{foot-1}. Usually, it is defined as the lengthening (shortening) of the wavelength (frequency) of light as it moves away from a massive body such as the Earth. Light is thought to loose energy as it reaches higher altitudes in a gravitational field due to its interaction with the field \cite{schild,MTW,feynman,schutz}. However, as it has been pointed out in \cite{synge} and emphasized in  \cite{okun}, light does not actually interact with the gravitational field and lose energy as it moves up to higher potentials. This is because light,  strictly speaking, does not have a gravitational mass and thereby cannot be assigned a potential energy. What happens in reality is that the frequency of light emitted by an atom (or a nucleus) at a lower gravitational potential is smaller than the frequency of light emitted by an identical atom/nucleus at a higher potential \cite{okun}.

Certain similarities between classical gravitational and electric fields lead one naturally to question if a similar effect might take place in a static electric field. Long before the experimental confirmation of the gravitational effect, this possibility was exercised and found that light does not undergo an actual frequency change as it moves in a static electric field from a low potential to a higher one or vice versa \cite{kennedy}. An interferometer was used to compare the frequencies of light before and after traveling in a static electric field. A second experiment ended up with the same null conclusion \cite{drill}. There was no attempt in these works to define and discuss the \emph{electrostatic redshift of light}. An endeavor to explain the null results of these experiments theoretically was presented in \cite{wood} by employing the Reissner-Nordstr\"om metric in general relativity.

The purpose of the present work is to discuss, to our knowledge for the first time in the literature, the effects that occur in a static electric field similar to those of the gravitational ones. To this end, we first expound the gravitational effects and then extend them to their electrical analogues. We present a general relativistic scheme that predicts the electrical effects too. We demonstrate that the electrostatic effects can also be obtained from an {\emph{electrical equivalence principle}. We propose an experiment with two $^{27}Al^+$ ions at different potentials in an electric field to measure the electrostatic time dilation effect. Finally, we argue that the Reissner - Nordstr\"om - treatment of the electrostatic redshift in \cite{wood}  to explain the null results of the electrical redshift experiments reported in \cite{kennedy, drill} involves a conceptual error and is irrelevant to electrical redshift.

\section{Review of the gravitational effects}

Let us consider an atom of mass $m$ at a height $H_0$ in a static gravitational field like that of the Earth. Taking this height as the reference level for the gravitational potential, the energy of the atom is

\be%1
E(0) = m_ic^2,
\label{eqn:groundE-0}
\ee
in its ground state, and
\be%2
E^*(0) = m_ic^2+\mathcal{E}=m_i^*c^2,
\label{eqn:excitedE-0}
\ee
in an excited state, denoted by a superscript ${^*}$, whose energy exceeds that of the ground state by $\mathcal{E}$.
Raising the atom by a height $H$ changes the energy levels to
\be %3
E(H)=m_ic^2+m_pgH,
\label{eqn:groundE-H}
\ee
in the ground state, and to
\be %4
E^*(H)=m_i^*c^2+m_p^*gH,
\label{eqn:excitedE-H}
\ee
in the excited state, where $g\approx 9.80\, m/s^2$ is the local value of the gravitational acceleration on the surface of the Earth, $m_i$ and $m_p$ are the inertial and passive gravitational masses. Here $m_i^*$ and $m^*_p$ are  the excited masses of the atom and are given by
\bear %5
m_i^*=m_i+\frac{\mathcal{E}}{c^2}.\nonumber\\
m_p^*=m_p+\frac{\mathcal{E}}{c^2}.
\eear
 As a result of this raising, the oscillation frequency $f=E/h$ of the atom changes from $f(0)$ to $f(H)$ in the ground state, and from $f^*(0)$ to $f^*(H)$ in the excited state, where $h$ is the Planck constant. These frequencies at different altitudes are related to each other in the ground and excited states, respectively as
\bear%6
f(H)=f(0)\left(1+\frac{m_p}{m_i}\frac{gH}{c^2}\right)=f(0)\left(1+\frac{gH}{c^2}\right),\nonumber\\
f^*(H)=f^*(0)\left(1+\frac{m^*_p}{m^*_i}\frac{gH}{c^2}\right)=f^*(0)\left(1+\frac{gH}{c^2}\right),
\eear
\noindent where we have used $m_p/m_i=m^*_p/m^*_i=1$. Therefore, the fractional changes in the oscillation frequencies due to this raising are given by
\bear%7
\frac{\delta f}{f(0)}=\frac{f(H)-f(0)}{f(0)}=\frac{gH}{c^2},\nonumber\\
\frac{\delta f^*}{f^*(0)}=\frac{f^*(H)-f^*(0)}{f^*(0)}=\frac{gH}{c^2},
\label{eqn:del-f}
\eear
which are the same both in the ground and excited states. Notice that $gH$ in these equations and what follows is equal to the change $\Delta\phi_g=\phi_g(H)-\phi_g(0)$, where $\phi_g$ is the gravitational potential. We note that these fractional changes are equal to the ratio of the work done against gravity by the external agent in raising the masses $m$ and $m^*$ by $H$, to their rest energies. Taken as a frequency reference, such an atom can be considered as a clock whose rate, its oscillation frequency, is faster the higher the height $H$ is. Thus, the time intervals measured by such a clock are proportional to its oscillation frequencies and are given by \cite{foot-2}
\bear %8
\Delta T(H)=\Delta T(0)\left(1+\frac{m_p}{m_i}\frac{gH}{c^2}\right)=\Delta T(0)\left(1+\frac{gH}{c^2}\right),\nonumber\\
\Delta T^*(H)=\Delta T^*(0)\left(1+\frac{m^*_p}{m^*_i}\frac{gH}{c^2}\right)=\Delta T^*(0)\left(1+\frac{gH}{c^2}\right),
\label{eqn:deltaT}
\eear
which indicate how such atoms/clocks age. Thus we find out that  atoms at higher altitudes, or higher potentials, age faster than those at lower altitudes from the gravitating source by a fractional change
\bear %9
\frac{\delta \Delta T}{\Delta T(0)}=\frac{\delta \Delta T^*}{\Delta T^*(0)}=\frac{gH}{c^2}=\frac{\Delta\phi_g}{c^2}.
\eear

As we will now show, the same results can be obtained by considering the photon emissions in the excited states. An excited atom at height $0$ whose energy is as in Eq.(\ref{eqn:excitedE-0}) may make a transition to the ground state by emitting a photon of frequency
\be %10
\nu(0)=\frac{E^*(0)-E(0)}{h}=\frac{\mathcal{E}}{h}.
\ee
\noindent When this photon is directed up, it cannot be absorbed by an atom at height $H$ in its ground state whose energy is as in
Eq.(\ref{eqn:groundE-H}) due to the energy deficiency. An energy of $\mathcal{E}+\mathcal{E}gH/c^2$ is required of the photon to excite the atom so that its energy has the value in Eq.(\ref{eqn:excitedE-H}). As has been emphasized in \cite{okun}, the photon with energy $\mathcal{E}$ does not loose energy as it moves up in the gravitational field. In the experiment \cite{p-r,p-s}, the extra energy $\mathcal{E}gH/c^2$ the photon needs to have so as to be absorbed is supplied to it through the Doppler effect. The source, the iron - 57 nucleus at height $0$, is  moved up toward the absorber to increase  the energy of the photon to
\bear %11
\mathcal{E_{\gamma}}&=&\mathcal{E}\left(\frac{1+\frac{v}{c}}{1-\frac{v}{c}}\right)^{1/2}\nonumber\\
&\approx& \mathcal{E}+\mathcal{E}\frac{v}{c}.
\eear
This is achieved by adjusting the upward speed of the source, as in \cite{p-r,p-s}, to
\be %12
v=\frac{gH}{c}.
\label{eqn:speed}
\ee
As is pointed out in \cite{okun}, had the photon lost energy as it moved up, as is assumed in the wrong interpretation of the gravitational redshift, the required Doppler speed would have been twice as that in Eq.(\ref{eqn:speed}) \cite{foot-3}. The excited nucleus at $H$ with enegy $E^*(H)$ may then make a transition to its ground state. The frequency of the photon emitted in this transition is
\bear %13
\nu(H)&=&\frac{E^*(H)-E(H)}{h}=\frac{\mathcal{E}_0+\mathcal{E}_0gH/c^2}{h}\nonumber\\
&=&\nu(0)\left(1+\frac{gH}{c^2}\right),
\eear
which is naturally equal to the frequency of the photon absorbed. The fractional change $\delta\nu / \nu(0)$ in the frequencies of the photons emitted and the predicted aging of the nuclei at different altitudes are, respectively,  the same as those in Eq.(\ref{eqn:del-f}) and Eq.(\ref{eqn:deltaT}).

One may wonder why the photon is still said to undergo a redshift even though its energy/frequency does not change as it moves up or down in a gravitational field. As is emphasized in \cite{okun}, this is because the energy levels of atoms/nuclei at higher altitudes undergo a blueshift, namely their energies increase, relative to atoms/nuclei at lower altitudes. The photon emitted by a nucleus/atom at a lower altitude and absorbed by an identical atom/nucleus at a higher altitude is seen by this atom/nucleus to be redshifted. This can be understood as follows: The higher-up-atom/nucleus, the clock, measures the frequency $\nu(0)$ of the photon when it reaches it as $\bar{\nu}(H)$, given by
\bear 
\bar{\nu}(H)&=&\bar{\nu}(0)\left(\frac{\nu(0)}{\nu(H)}\right)\nonumber\\
&=&\nu(0)\left(1-\frac{gH}{c^2}\right),
\label{eqn:nu-bar}
\eear
with a fractional change of $-gH/c^2$, and $\bar{\nu}(0)=\nu(0)$.

To sum up what we have reviewed so far, our analysis shows clearly that (i) the gravitational time dilation effect is due to the difference in the total energies of objects when they are at different altitudes in a gravitational field, (ii) the frequency of photons do not change as they move in a gravitational field, and (iii) atoms/nuclei higher up in a gravitational field whose energies are blushifted see the photons emitted by atoms/nuclei lower down in the field as redshifted.

\section{Gravitational time dilation from general relativity}

As is well known, the gravitational time dilation effect can be obtained from the Schwarzschild line element
\be 
ds^2=-\left(1-\frac{2GM}{c^2r}\right)c^2dt^2+\left(1-\frac{2GM}{c^2r}\right)^{-1}dr^2
+r^2d\theta^2+r^2sin^2\theta d\phi^2,
\ee
where $G$ is the gravitational constant,  $M$  the mass of a spherical body, and $c$ the speed of light. The proper time interval $d\tau$ for a test particle located at the coordinate $r$ is related to its
coordinate time interval $dt$ by the relation
\be %16
d\tau=\left(1-\frac{2GM}{c^2r}\right)^{1/2}dt.
\ee
The ratio of the proper time intervals for two clocks (test particles) at $r=R+H$ and $r=R$ is thus
\be %17
\frac{d\tau(R+H)}{d\tau(R)}=\frac{\Delta T(H)}{\Delta T(0)}=\left(1-\frac{2GM}{c^2(R+H)}\right)^{1/2}\Big{/}\left(1-\frac{2GM}{c^2R}\right)^{1/2},
\ee
because the two coordinate time intervals are equal. Since the height $H$ of the top clock is much smaller than the radius $R$ of the Earth and the terms with $c^{-2}$ are much smaller than $1$ the above ratio is approximated by
\be %18
\frac{\Delta T(H)}{\Delta T(0)}\approx\left(1+\frac{gH}{c^2}\right),
\ee
where $g=GM/R^2$. This is the same expression as that in Eq.(\ref{eqn:deltaT}). Thus the gravitational time dilation is indeed a prediction of the Schwarzschild line element.

\section{Electrostatic time dilation and redshift}
\label{sec:electro} 
Next, we undertake the electrical analogues, if they exist, of the gravitational effects. To this end, let us consider an electric field ${\bf E}$ in a region of the $xy$ plane directed from right to left in the $-x$ axis. Assume there are four identical positively charged atoms (cations) in their ground states at a level on the $x$ axis where the electrical potential  will be taken to be zero. The atoms  are prevented from interacting with each other through some mechanism. The energies of the atoms at this zero level of potential are given by
Eq.(\ref{eqn:groundE-0}), as in the gravitational case. Then we move one of the atoms to the right  by a distance $d$ on the $x$ axis. The energy of this atom is \cite{foot-4}
\be %19
E(d)=m_ic^2+q|{\bf E}| d,
\label{eqn:elecE-d}
\ee
where $q$ is the positive charge of the atom. Then, we excite two of the atoms by letting them absorb  a photon of energy $\mathcal{E}$ increasing their energy to
\be %20
E^*(0)=m_ic^2+\mathcal{E}.
\label{eqn:exctdE-0}
\ee
\noindent Afterwards, we move one of these excited atoms to the right by the same distance $d$ as before. The energy of this atom will be

\be %21
E^*(d)=m_ic^2+\mathcal{E}+ q|{\bf E}| d.
\ee
\noindent The fractional changes in the energies and vibrational frequencies are
\be %22
\frac{\delta E}{E(0)}= \frac{\delta f}{f(0)}=\frac{q}{m_i}\frac{|{\bf E}| d}{c^2}=\frac{q}{m_i}\frac{\Delta\phi_e}{c^2},
\label{eqn:frac-f}
\ee
in the ground state, and
\be %23
\frac{\delta E^*}{E^*(0)}= \frac{\delta f^*}{f^*(0)}=\frac{q}{m_i}\frac{|{\bf E}| d}{c^2\left(1+\frac{\mathcal{E}}{ m_ic^2}\right)}\approx
\frac{q}{m_i}\frac{|{\bf E}| d}{c^2}=\frac{q}{m_i}\frac{\Delta\phi_e}{c^2},
\ee
in the excited state where $\mathcal{E}/mc^2\ll 1$ has been implemented. Here \mbox{$\Delta\phi_e=|{\bf E}| d>0$} is the electrical potential difference between the positions of the two atoms. The lessons learned in our discussion of the gravitational case lead us to conclude that the aging of these atoms  takes place according to
\bear %24
\Delta T(d)=\Delta T(0) \left(1+\frac{q}{m_i}\frac{|{\bf E}| d}{c^2}\right),\nonumber\\
\Delta T^*(d)=\Delta T^*(0) \left(1+\frac{q}{m_i}\frac{|{\bf E}| d}{c^2}\right),
\label{eqn:delta-T-E}
\eear
which are similar to the gravitational ones in Eq.(\ref{eqn:deltaT}). It should be noted that these electrical time dilation effects become contraction effects for anions, atoms with a total negative electric charge, for \mbox{$\Delta\phi_e>0$}. This is an effect with no counterpart in the gravitational case due to the absence of negative  mass. The excited atoms at $x=0$ and at $x=d$ can make a transition to their ground states by emitting a photon of frequency
\bear %25
\nu(0)=\frac{E^*(0)-E(0)}{h}=\frac{\mathcal{E}}{h},\nonumber\\
\nu(d)=\frac{E^*(d)-E(d)}{h}=\frac{\mathcal{E}}{h}.
\label{eq:frequency}
\eear
It is no surprise that these frequencies are equal. This is a consequence of the conservation of electric charge. Moving an electric charge from one point to another in an electric field by a distance $d$ does not change the amount of the electric charge whereas moving it vertically by a height $H$ in a gravitational field would change the gravitational mass by $gH/c^2$.

Let us contemplate a Pound - Rebka - Snider experiment \cite{p-r, p-s} performed in an electric field. The photon of energy $\mathcal{E}$ emitted by the source/emitter at $x=0$, whose energy is as given in Eq.(\ref{eqn:exctdE-0}),  will reach the absorber/receiver atom, which is in its ground state at $x=d$ whose energy is as given in Eq.(\ref{eqn:elecE-d}). The photon can be absorbed by this atom without the need to increase its energy by a Doppler shift because the mass of the atom plays no role in the energy shift in an electric field in contrast with the gravitational case. Having absorbed the photon, this excited atom can in turn emit the photon with the same energy $\mathcal{E}$ by making a transition to its ground state. The overall result is null. There is no change in the frequencies of the absorbed and emitted photons. Thus no electrical redshift! This is also predicted by the electrical analogue of Eq.(\ref{eqn:nu-bar}), which would yield $\bar{\nu}(d)=\nu(0)$. We can now understand the null results of the electrical redshift experiments. In \cite{kennedy}, light of a certain frequency from an electrodeless discharge in mercury vapor was let to travel through a potential difference of $50 kV$ above or below ground. A quartz interferometer kept at zero potential was used to measure the frequency of the received light. No change in the frequency of the light was detected. This experiment was repeated in \cite{drill} by employing a potential difference of $\pm 300 kV$ and the findings in \cite{kennedy} were improved by a factor of ten. Theoretically, these results are expected because the energy/frequency of light does not undergo a shift as it travels between different electrostatic potentials in an electric field, just as what happens in a gravitational field as we have discussed.
\section{Could there be a general relativistic explanation of the electrostatic time dilation?}
	
Comparing equations (\ref{eqn:deltaT})	and (\ref{eqn:delta-T-E}) reveals that the latter can be obtained from the former on replacing $m_p$ with $q$, $g$ with $|{\bf E|}$ and $H$ with $d$. This could be interpreted as the possibility that there might exist a general relativistic theory of electromagnetism which is very similar to that of gravitation. This theory might be a unified theory of gravitation and electromagnetism. It might predict \cite{ozer} the spacetime metric outside a spherical body of mass $M$ and electric charge $Q$ as
\begin{eqnarray} 
ds^2=-\left(1-\frac{m_p}{m_i}\frac{2GM}{c^2r}+\frac{q}{m_i}\frac{2k_eQ}{c^2r}\right)c^2dt^2+\left(1-\frac{m_p}{m_i}\frac{2GM}{c^2r}+\frac{q}{m_i}\frac{2k_eQ}{c^2r}\right)^{-1}dr^2
+\nonumber\\
r^2d\theta^2+r^2sin^2\theta d\phi^2,
\label{eqn:uni-metric}
\end{eqnarray}
where $k_e$ is the Coulomb constant and $m_p/m_i=1$. It just follows that the proper and coordinate time intervals of a test particle located at $r$ are related as
\be 
d\tau=\left(1-\frac{2GM}{c^2r}+\frac{q}{m_i}\frac{2k_eQ}{c^2r}\right)^{1/2}dt.
\ee

Let us contemplate an experiment with a charged metallic sphere with a radius $R$ of few decimeters. Assume that the sphere is negatively charged ( $Q=-|Q|$ ) so that its electric field is directed from the right to the left along the horizontal towards the equator of the sphere.
The ratio of the proper time intervals for two clocks (test particles) located on the equatorial plane at $r+d$ and $r$ for $r>R$ is thus
\bear %28
\frac{d\tau(r+d)}{d\tau(r)}=\frac{\Delta T(d)}{\Delta T(0)}=\left(1-2\frac{q}{m_i}\frac{k_e|Q|}{c^2(r+d)}\right)^{1/2}\Big{/}\left(1-2\frac{q}{m_i}\frac{k_e|Q|}{c^2r}\right)^{1/2}\nonumber\\,
=\left(1+2\frac{q}{m_ic^2}\phi_e(r+d)\right)^{1/2}\Big{/}\left(1+2\frac{q}{m_ic^2}\phi_e(r)\right)^{1/2}
\eear
because the two coordinate time intervals are equal and $\phi_e(r)=-k_e|Q|/r$ is the electric potential. There is no gravitational contribution due to the Earth   because the clocks are at the same height. The gravitational contribution of the sphere due to its mass is negligibly smaller than its electrical contribution.  Since the rest energy $m_ic^2$ of the test particle is much larger than its potential energy $q\phi_e$   the above ratio is approximated by
\be %29
\frac{\Delta T(d)}{\Delta T(0)}\approx\left(1+\frac{q}{m_ic^2}\Delta\phi_e\right)=\left(1+\frac{q}{m_i}\frac{| {\bf E}| d}{c^2}\right)
\label{frac-metric}
\ee
where $\Delta\phi_e=\phi_e(r+d)-\phi_e(r)>0$ is the electric potential difference between two points apart by a distance $d$ in the electric field ${\bf E}$. This is the same expression as the first one  in Eq.(\ref{eqn:delta-T-E}). It should be noted that even though the electric field and the potential difference in Eq.(\ref{frac-metric}) are due to a charged metal sphere, the result is general as the considerations in section \ref{sec:electro} imply. The potential difference may be that in an electric field created by other means, as in section \ref{sec:proposed}. Thus the metric in Eq.(\ref{eqn:uni-metric}) provides a complete explanation both for the gravitational and electrostatic time dilation effects. The exact one - to - one similarity of the electrostatic time dilation to its gravitational counterpart both in the classical energy considerations and the general relativistic treatments and the fact that the gravitational one is a fact of Nature suffice it to conclude that the electrostatic effect, too, must be a fact of Nature.
\section{Could there be an electrical equivalence principle from which the electrostatic time dilation ensues?}
\label{sec:elec-EP}
The answer to this question is affirmative. Electrostatic experiments performed in an accelerated frame would produce identical results as electrostatic experiments, with a single charged particle or particles of the same type, performed in a uniform electric field. To this end, let us consider a cabin in a rocket in deep space where there are no fields of any kind. Let there be two charged particles (observers) of the same type at the bottom and the top of the cabin of height $d$. Let us establish a Cartesian coordinate frame and choose the direction of motion of the rocket as the upward z-axis. Let the bottom of the cabin and the origin of the coordinate frame overlap at time $t=0$. Let the rocket, hence the cabin, accelerate upward at ${\bf a}_{cabin}=-\frac{q}{m_i}{\bf E}$ , where $q$ is the electric charge of some known particle, $m_i$ is its inertial mass, and ${\bf E}$ is some known electric field directed in the negative $z$ direction as the gravitational field for convenience. The charge $q$ will be taken positive in the following discussion for simplicity. Accordingly, adapting the gravitational equations in \cite{hartle} to electricity (and reversing the direction of the light pulses emitted) the positions of the bottom and top charges (observers) will be given by
\begin{align}
z_B(t)&=\frac{1}{2}\frac{q}{m_i}|{\bf E}|t^2,\nonumber\\
z_T(t)&=d+\frac{1}{2}\frac{q}{m_i}|{\bf E}|t^2.
\end{align}
The charge (observer) at the bottom emits  two light pulses directed upward at $t=t_1$ and at $t=t_2=t_1+\Delta\tau_B$ seperated by a time interval of $\Delta\tau_B$. Let these pulses be received by the charge at the top at times $t=t_1^`$ and $t=t_2^`=t_1`+\Delta\tau_T$ separated by a time interval of $\Delta\tau_T$.The distance traveled by the first pulse is
\be
z_T(t_1^`)-z_B(t_1)=c(t_1`-t_1),
\ee
which gives
\be
d+\frac{1}{2}\frac{q}{m_i}|{\bf E}|{t_1^`}^2-\frac{1}{2}\frac{q}{m_i}|{\bf E}|t_1^2=c(t_1^`-t_1).
\label{eq:el-ep-1}
\ee%
Similarly, the distance traveled by the second pulse is
\be 
z_T(t_1^`+\Delta\tau_T)-z_B(t_1+\Delta\tau_B)=c(t_1`+\Delta\tau_T-t_1-\Delta\tau_B), 
\ee 
which gives 
\be
d+\frac{1}{2}\frac{q}{m_i}|{\bf E}|{(t_1^`+\Delta\tau_T})^2-\frac{1}{2}\frac{q}{m_i}|{\bf E}|{(t_1+\Delta\tau_B})^2=c(t_1`+\Delta\tau_T-t_1-\Delta\tau_B), 
\label{eq:el-ep-2}
\ee
Using Eq.(\ref{eq:el-ep-1}) in Eq.(\ref{eq:el-ep-2}) and setting $t_1^`=t_1+d/c$ and neglecting all terms second  order in time, we get
\be
\Delta\tau_T(c-\frac{q}{m_i}|{\bf E}|\frac{d}{c})\approx c\Delta\tau_B
\ee
or,
\be
\Delta\tau_T\approx \Delta\tau_B(1+\frac{q}{m_i}|{\bf E}|\frac{d}{c^2}).
\label{eq:el-ep-3}
\ee 
This is the same  as  Eq.(\ref{eqn:delta-T-E}) above \cite{foot-5}. Now, the electrical equivalence principle tells us that this accelerated frame in deep space where there exists no fields of any kind is equivalent to a stationary frame (a lab) on Earth where there exists a uniform downward electric field  and the elapsed times between two emissions of photons by particles (atoms, etc.) of charge $q$  seperated by a distance $d$ is given by Eq.(\ref{eq:el-ep-3}) above. It should be noted that the equivalence of the two frames is realized not for a unique value, but for all values of $q/m_i$  of the  particles of the same type with which the experiments are done. This point is of utmost importance that it deserves to be elucidated further.  

 To this end, consider a collection of charged particles  with different $q/m_i$ ratios moving  in an external electric field ${\bf E}$. Adapting the gravitational equations in \cite{wein} to electricity, the equation of motion of the $nth$ particle will be
%
% eq. 
\be \label{eqparticlesaccel} % 12
m _i^{(n)}\frac{d^2{\bf r}\,^{(n)}}{dt^2}=q^{(n)}{\bf E}+\sum_{M}{\bf F}( {\bf r}\,^{(n)}-{\bf r}\,^{(M)}),\hspace{0.5cm}n=1,2,....,N,
\ee
where  ${\bf F}$ denotes the interparticle interactions. The spacetime transformations
\be % 
t'=t,\hspace{0.5cm}{\bf r}\,^{'(n)}={\bf r}\,^{(n)}-\frac{1}{2}\,\left(q^{(n)}/m_i^{(n)}\right)\,{\bf E}\,t\,'^2,\hspace{0.5cm}n=1,2,....,N,
\label{eq:electrspacetimetrans}
\ee
cast Eq. (\ref{eqparticlesaccel}) to
\bear % 
\label{eqsümeyye}
m _i^{(n)}\frac{d^2{\bf r}\,^{'(n)}}{dt^{'2}}+m_i^{(n)}\left (q^{(n)}/m_i^{(n)}\right ){\bf E}&=&q^{(n)}{\bf E}+\sum_M{\bf F}({\bf r}\,^{'(n)} -{\bf r}\,^{'(M)})\nonumber\\
m _i^{(n)}\frac{d^2{\bf r}\,^{'(n)}}{dt^{'2}}&=&\sum_M{\bf F}({\bf r}\,^{'(n)} -{\bf r}\,^{'(M)}).
\eear
The electric force on the $nth$ particle
\be \label{eqinertialforce} % 16
{\bf F}_{E}^{(n)}= q^{(n)}{\bf E}
\ee
has been cancelled by the following fictitious force
\be \label{eqinertialforce1} % 
{\bf F}_{fict}^{(n)}=- m_i^{(n)}\left (q^{(n)}/m_i^{(n)}\right ){\bf E}=-q^{(n)}{\bf E}.
\ee
In other words,
\be  % 
{\bf F}_{E}^{(n)}+{\bf F}_{fict}^{(n)}=0.
\label{eq:E-fict}
\ee
As is seen clearly from Eq.(\ref{eqinertialforce1}) that there is no restriction on the ratio $(q^{(n)}/m_i^{(n)})$  for the cancelation of the external  electric force on the particle locally. Especially, it is not required for this cancellation  that this ratio  be $1$. It can be equal to any value observed in Nature. Furthermore, the cancelation of the electric field in the neighbohood of the $nth$ particle occurs independently of the cancelation of the electric field for the other particles. Thus  the electrical equivalence principle can be stated, adapting the gravitational statement in \cite{tollman} to electricity,   as "It is always possible at any space-time point of interest to transform to coordinates such that the effects of electricity will disappear over a differential region in the neighborhood of that point, which is taken small enough so that the spatial and temporal variation of electricity within the region may be neglected." It can also be stated in the form "It is impossible to distinguish the fictitious inertial force from the real electrical force in a local region containing a single particle".

We reiterate, just as the equivalence of an accelerating frame in deep space and a frame at rest in a local uniform gravitational  field, a similar equivalence exists for electricity too. A frame containing a charged particle of mass $m_i$ and charge $q$ and accelerating in deep space at an acceleration ${\bf a}=-(q/m_i){\bf E}$ is equivalent to a laboratory frame at rest where there is an identical particle in it and a uniform electric field ${\bf E}$. As has been demonstrated above, the electricity experiments involving the charged particle in these two frames  give identical results. As in gravity, there exists a second pair of equivalent frames that result in by the addition of an acceleration ${\bf a}=(q/m_i){\bf E}$ to the deep space - frame and the Earth - frame. Thus the fame accelerating upward is replaced by a frame at rest in deep space, irrespective of whether it is charged or not, and the frame at rest in an electric field on Earth is replaced by a charged frame with $Q/M=q/m_i$ falling with acceleration ${\bf a}=(q/m_i){\bf E}$ in a uniform electric field. These two frames are completely equivalent as far as the results of the experiments performed with the charged particle at hand.

The effect of applying the spacetime transformation in Eq.(\ref{eq:electrspacetimetrans}) on each particle is equivalent to putting each particle in a small enough cabin whose    $Q/M_i$ ratio is the same as the $q/m_i$ ratio of the particle in it. Each such cabin will be a local reference frame with Cartesian coordinates.
The $nth$  cabin for example, will be falling freely at an acceleration
${\bf a}^{(n)}=Q^{(n)}/M^{(n)}_i{\bf E}$. If the interparticle interactions are neglected, each cabin would be a local inertial frame.
The acceleration of the $n$th particle relative to its own hypothetical cabin as it falls will be \cite{ozer,matos}
\be \label{eqrelaccel} % 19
{\bf a}^{(n)}_{rel}\ =\left(\frac{q^{(n)}}{m_i^{(n)}}-\frac{Q^{(n)}}{M^{(n)}_i}\right){\bf E}=0.
\ee

\noindent Each such cabin with a charged particle in it is equivalent to a similar cabin floating in deep space.
Though there is no need, we can group the particles with the same $q/m_i$ ratio and put them in a larger cabin whose charge-to-mass ratio $Q/M$ is the same and falling freely in the electric field ${\bf E}$.  Obviously, it is not legitimate to put various charged particles with different $q/m_i$ ratios in a falling frame and require that they all float in this frame. This is because this would compel the $q/m_i$ ratios of the particles to be the same and equal to the $Q/M$ ratio of the falling cabin. Because any spacetime point contains a single particle only, the equivalence principle stated above may be called $\emph{ the single - particle equivalence principle}$ \cite{ozer,matos}.

\section{Proposed experiment to measure the electrostatic time dilation}
\label{sec:proposed}

Before we pass to the next item, we point out to the experimental possibility of measuring the electrostatic time dilation effect by high precision clocks similar to the atomic ones. What is required for such a measurement is $\emph{ionic clocks}$ \cite{ionic1,ionic2,ionic3} whose oscillators are positively (or, in principle, negatively) charged ionic atoms whose energy levels will split in an electric field depending on their positions in the field \cite{foot-4}. For example, the fractional change in the oscillation frequencies of two $^{27}Al^+$ optical ion clocks in a static electric field for a potential difference of
$\Delta\phi_e(in V)$ between them would be
\be %30
\frac{\delta f}{f(0)}=\frac{q}{m}\frac{\Delta\phi_e}{c^2}=3.979\times 10^{-11}\Delta\phi_e/V,
\ee
where $q=e=1.602 \times 10^{-16}C$ is the charge of $^{27}Al^+$ and $m=4.480\times 10^{-26}kg$ its mass. This would be much larger than the special relativistic and gravitational time dilation effects reported in \cite{chou} depending on the value of
$\Delta\phi_e$.
The experimental confirmation of the electrostatic time dilation effect would be an unequivocal indication for the correctness of the unification metric in Eq.(\ref{eqn:uni-metric}).
\section{The Reissner - N\"ordstrom metric and the electrostatic redshift}

Finally, we comment on the work in \cite{wood}, where an attempt to explain the null results of the electrostatic redshift experiments reported in \cite{kennedy, drill} was made, by reproducing the derivation of the general relativistic prediction of the electrical redshift in \cite{wood}. They start off by giving the fractional change in the frequency of light in general relativity, which is
\be %31
\Delta\nu/\nu=1-\left(g_{00}/g'_{00}\right)^{1/2},
\label{eqn:delta-RN}
\ee
where $\Delta\nu=\nu_{observed}-\nu$ with $\nu$ being the frequency of the light emitted by the source. $g_{00}$ is the coefficient of the timelike coordinate in the square of the differential line element
\be %32
ds^2=g_{\mu\nu}dx^{\nu}dx^{\nu}
\ee
at the position of the absorver/receiver, and $g'_{00}$ being the one at the position of  the source/emitter. The absorber is assumed to be located on the dome of an electrostatic accelerator (like a van de Graaf generator) whose electrostatic potential is given by (in SI units)
\be %33
\varphi=k_eQ/R,
\ee
where  $Q$ is the charge of the dome, $R$  the radius of the dome. The $g_{00}$ on the dome is assumed to be that of the
Reissner - Nordstr\"om line element given by
\be %34
g_{00}=1-2GM/Rc^2+k_eGQ^2/R^2c^4,
\label{eqn:g00-RN}
\ee
where $M$ is the mass of the dome. The authors say they are only interested in the electrostatic effect and thereby drop the second term in the equation above as a result of which Eq.(\ref{eqn:delta-RN}) becomes
\be %35
\Delta\nu/\nu=1-\left(1+G\varphi^2/k_ec^4\right)^{1/2},
\ee
where $g'_{00}=1$ because the source is assumed to be in a region of zero electrostatic  potential inside the dome. Since the potential term is much smaller than one, this reduces to
\be %36
\Delta\nu/\nu=-G\varphi^2/2k_ec^4.
\ee
\noindent For an electrostatic accelerator of several $MeV$, $\Delta\nu/\nu\sim 10^{-40}$, and the null results in the experiments \cite{kennedy,drill} are expected, as proclaimed by the authors. The conceptual error in this treatment is that the Reissner - Nordstr\"om metric is a solution to the Einstein - Maxwell field equations for an electrically charged spherical body of mass $M$ and charge $Q$. The third term on the right of
Eq.(\ref{eqn:g00-RN}) is not an electrostatic term, rather it is a gravitational one just like the second term. The meaning of the third term is as follows: The electric field  outside the central body due to its charge $Q$ has an energy $U_{field}\sim r^3\epsilon_0| {\bf E}| ^2/2$ which is equivalent to a mass $M_{field}=U_{field}/c^2\sim k_eQ^2/c^2r$. This mass gives rise to a gravitational potential $\Phi_g =GM_{field}/r$ whose contribution to $g_{00}$ must be proportional to $\Phi_g/c^2$ which is approximately equal to $Gk_eQ^2/c^4R^2$ for $r=R$. This is the third term in $g_{00}$ in Eq.(\ref{eqn:g00-RN}). Thus on the surface of the dome $g_{00}$ can be written as
\be %37
g_{00}=1-2GM_{eff}(R)/Rc^2.
\ee
with $M_{eff}(r)$ being the effective mass of the dome, the central body, given by
\be %38
M_{eff}(r)=M-\frac{1}{2}\frac{k_eQ^2}{rc^2},
\ee
where $r\ge R$ is the radial coordinate. Therefore, according to the meaning of the Reissner - Nordstr\"om metric, light moving in this spacetime is moving in a gravitational field created by the mass $M_{eff}$, and the redshift it may undergo is a gravitational one. Certainly, it is incorrect to call the contribution of the $Q^2$ - term an electrostatic redshift.

\section{Conclusions}
In the present work, we have shown, employing classical electromagnetic theory, that positively  charged objects age faster at high  potential points than those at low potential points in an electric field. The aging of the negatively charged objects, however, takes place in the opposite way. This electrostatic time dilation effect is similar to its gravitational analogue and there exists a general relativistic theory  that predicts its existence \cite{ozer}. We have  demonstrated that this effect can also be obtained from an electrical equivalence principle whose salient features have been elucidated.   A
Pound - Rebka - Snider experiment subjecting the source to a horizontal electric field would show no electrical redshift because electric charge is conserved and is independent of any forms of energy (which is not the case for mass). This is a result confirmed by two
experiments that employed interferometers to observe  the questioned electrostatic redshift. If, on the other hand, it were possible to measure the frequency change of photons in the gravitational field of the Earth with interferometers, one would observe no gravitational redshift either. This is because the energy/frequency of photons do not change as they move in a gravitational field, as discussed in \cite{okun}. An attempt in the literature to explain the null results of the two electrostatic redshift experiments by making use of the Reissner - Nordstr\"om metric has been shown to be invalid. We cannot overemphasize the performance of the proposed  experiment with ionic clocks in a static electric field to confirm the electrostatic time dilation effect presented in this work. Such a confirmation would also support the unified theory of gravitation and electromagnetism leading to the metric in Eq.(\ref{eqn:uni-metric}).\\

\section*{Acknowledgement}
We thank C. J. de Matos of Institute of Aerospace Engineering, Technische Universit\"at Dresden, Germany  for invaluable discussions and bringing references \cite{galileo-1}, \cite{galileo-2}, and \cite{chou}  to our attention.


\begin{thebibliography}{}
\bibitem{foot-2} It should be noted that $\Delta T(H)$ or $\Delta T^*(H)$ are related to the oscillation frequencies as $\Delta T(H)/\Delta T(0)=f(H)/f(0)$, but not as $\Delta T(H)/\Delta T(0)=f(0)/f(H)$ as one may expect. This is because the ticks of clocks at height $0$ may be normalized so that the time intervel between two ticks is one second. Then clocks at height $H$ register a time interval of $N(H)$ seconds for $N(H)+1$ ticks while those at height $0$ register $N(0)$ seconds for $N(0)+1$ ticks. Obviously $(N(H)+1)/(N(0)+1)=f(H)/f(0)$.
\bibitem{foot-3} This is because, had it lost energy in moving up in the gravitational field, the photon of energy $\mathcal{E}$ at level $0$ would have reached height $H$ with an energy $\mathcal{E}-\mathcal{E}gH/c^2$. Hence the required Doppler speed would have been twice as that in Eq.(\ref{eqn:speed}) so that the energy of the photon is  $\mathcal{E}+\mathcal{E}gH/c^2$, neglecting the $c^{-4}$ and smaller terms.
\bibitem{foot-4} This should not be confused with the Stark Effect which takes place in an electric field as a result of the interaction of the field with the electric dipole moment of the atom. The effect considered here takes place  for electrically charged systems with zero elecric dipole moment too, such as a proton, for example. Furthermore, two charged atoms with the same electric dipole moment in a  uniform electric field but at different potentials would have the same Stark effect energy level splittings whereas their aging would be different due to the different electric potentials they are exposed to.
\bibitem{foot-5} Naively thinking, one may be tempted to conclude from this expression that the photons emitted at the bottom with frequency $\nu_B$ reach the top with frequency $\nu_T$ such that $\nu_T\approx\nu_B\left(1-\frac{q}{m_i}|{\bf E}|\frac{d}{c^2}\right)$. This is outright wrong! As has been discussed after Eq.(\ref{eq:frequency}), the frequency of the photons do not change as they travel in an electric field.
\end{thebibliography}
\end{document}